\documentclass[11pt,twoside]{article}

\usepackage{graphicx}
\usepackage{wrapfig}
\usepackage{cite}

\newcommand{\bc}{\begin{center}}
\newcommand{\ec}{\end{center}}
\begin{document}
\noindent\textit{\Large
\tabcolsep=0mm
\begin{tabular}{c}
\rule{175mm}{0mm}\\[-6mm]%\\[-1mm]
LIGHT-PRESSURE EXPERIMENTS BY P. N.~LEBEDEV \\[-1mm]
AND MODERN PROBLEMS \\[-1mm]
OF OPTOMECHANICS AND QUANTUM OPTICS
\end{tabular}}

\bigskip

\bc
{\bf Valentina M. Berezanskaya, Igor Ya. Doskoch, and Margarita A.
Man'ko$^*$}

\medskip

{\it
Lebedev Physical Institute, Russian Academy of Sciences\\
Leninskii Prospect 53, Moscow 119991, Russia}
\medskip

$^*$Corresponding author e-mail:~~~mmanko\,@\,sci.lebedev.ru
\ec

\begin{abstract}\noindent
In connection with the 150$^{\rm th}$ Anniversary of P.~N.~Lebedev,
we present historical aspects of his scientific and organizing
activity and recall his famous experimental observations and proof
of the existence of light pressure along with other results that
essentially influenced the development of physics in Russia and in
the whole world as well. We discuss the relationship of these
studies of electromagnetic waves and other kinds of vibrational
phenomena investigated by P.~N.~Lebedev to modern studies of the
interaction of photons with mirrors, gravitational waves, acoustic
waves, nonstationary (dynamical) Casimir effect of photon creation
in resonators with vibrating boundaries, and vibrations of voltage
and current in superconducting  circuits realizing the states of
qubits and qudits. We discuss the possibility of existence of the
nonstationary Casimir effect for gravitational waves and sound in
liquid helium.

\end{abstract}

\medskip

\noindent{\bf Keywords:} history of physics, P.~N.~Lebedev, light
pressure, nonstationary Casimir effect, superconducting circuit,
gravitational waves, sound in hellium.

\section{Introduction}
From the very beginning of studying the properties of light and the
applications of optical instruments, e.g., in astronomy, the nature
of light and the understanding of the essence of optical phenomena
has attracted the attention of researchers. Today, the results of
such researches reflecting the efforts of investigations of many
centuries in different countries can be formulated using two
different pictures. One picture corresponds to the so-called
corpuscular theory of light where the light propagation is
considered as a collection of moving particles. The other picture
corresponds to the propagation of waves either in the empty space or
in the media, and the waves demonstrate all phenomena known for the
waves such as diffraction and interference confirmed by many optical
experiments.

On the other hand, an important ingredient in understanding the
properties of light is to explain how the light interacts with
matter. Since the light can be reflected by a mirror, this
reflection is associated with the light action on the material of
the mirror. From the corpuscular picture of the light propagation,
it is quite clear that the particles kicking the mirrors and being
reflected by the mirror must press on this mirror. Nevertheless,
even such a clear picture in physics should be proved
experimentally. The name of Prof.~P.~N.~Lebedev, whose the 150 Years
Jubilee we are celebrating, is connected with this fundamental
problem.

In this paper, we review the Lebedev experiment described in
\cite{AnnPhys-1901}. The other goal of this paper is to discuss some
modern problems of quantum optics and optomechanics connected with
recent new experiments such as the observation of photons created
from the vacuum in resonators with moving boundaries based on
Josephson-junction superconducting devices~\cite{Sweden} and the
observation of gravitational waves~\cite{Science2016}. Also we
discuss the other wave phenomena studied by P.~N.~Lebedev such as
the experiments with waves on the water surface. The acoustic waves
related to the sound propagation in media were the subject of his
research as well. The organization of the work of this group of
physicists, who were pupils of P.~N.~Lebedev, and the creation of
the laboratory which grew later to become the Physical Institute,
will also be discussed in this paper. We recall P.~N.~Lebedev's
project of gathering the scientists and creating the scientific
laboratory, which grew fast and later on was transformed to the
Lebedev Physical Institute.

\section{Light Pressure}

\begin{wrapfigure}[24]{r}{76mm}
\vspace{-4mm}
\centerline{\includegraphics[width=76mm]{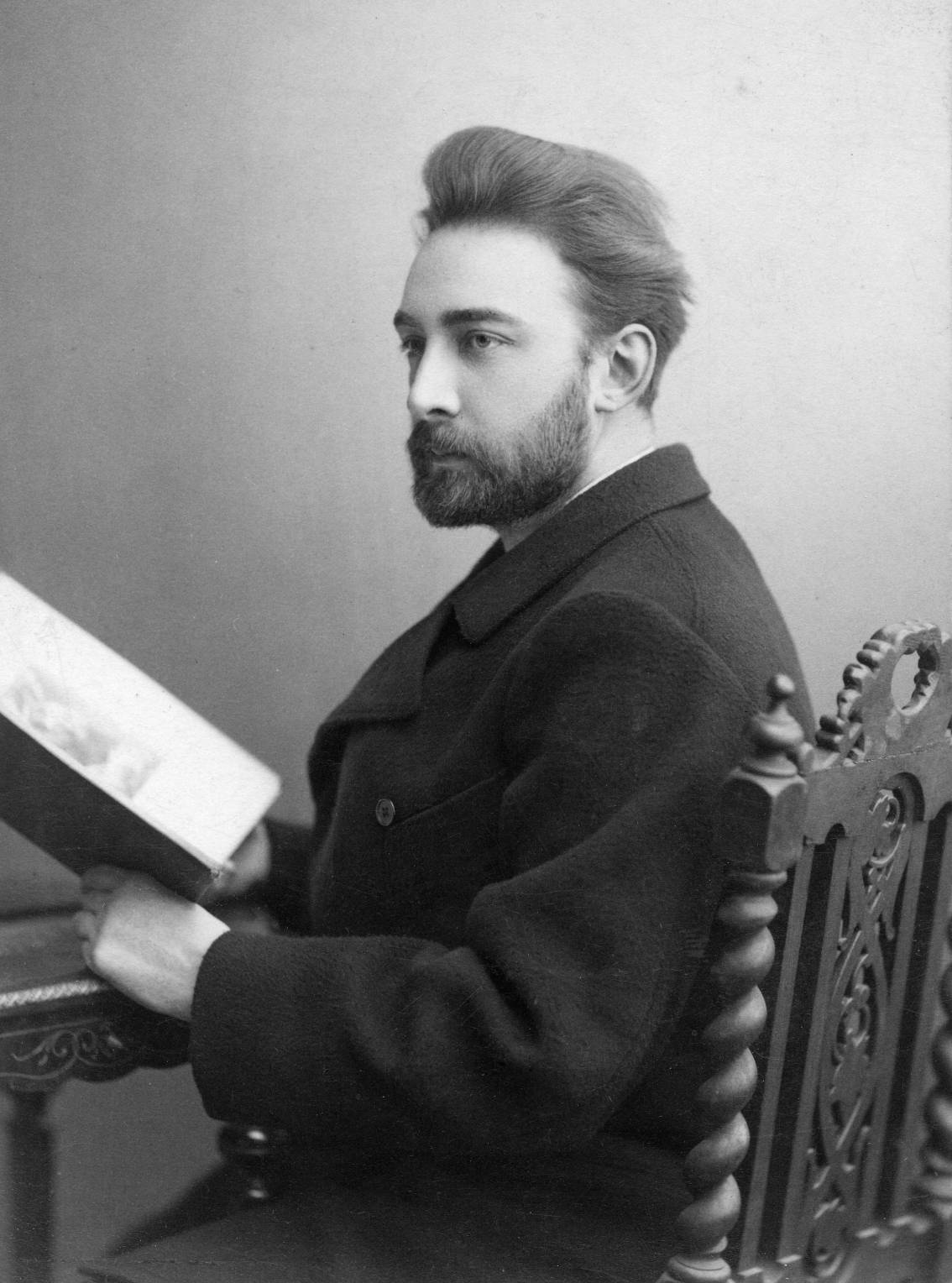}}
\vspace{2mm}
\centerline{\small P. N. Lebedev}
\end{wrapfigure}

The central result of P.~N.Lebedev was experimental observation of
the light pressure published in Annalen der Physik in
1901~\cite{AnnPhys-1901}. The existence of the light pressure
directly follows from the Maxwell equations, which have solutions
describing the electromagnetic waves. These waves are associated
with vibrations of the electric and magnetic fields propagating in
vacuum with the light velocity $c=299,792,458$~m/s (or $c\approx
300,000$~km/s). The Maxwell equations connect vibrations of the
magnetic and electric fields in an explicit form, and these
vibrations carry the energy. The properties of light were discussed
in our paper~\cite{Doskoch} dedicated to the International Year of
Light and Light-based Technologies announced by the UN General
Assembly for 2015, where historical aspects of the study of light
performed during several centuries along with modern results in the
area of both fundamental research of quantum phenomena in photon
systems and applications of laser technologies were presented.

\begin{figure}[ht]
%\begin{wrapfigure}[16]{r}{123mm}
\vspace{-1mm}
\centerline{\includegraphics[width=172mm]{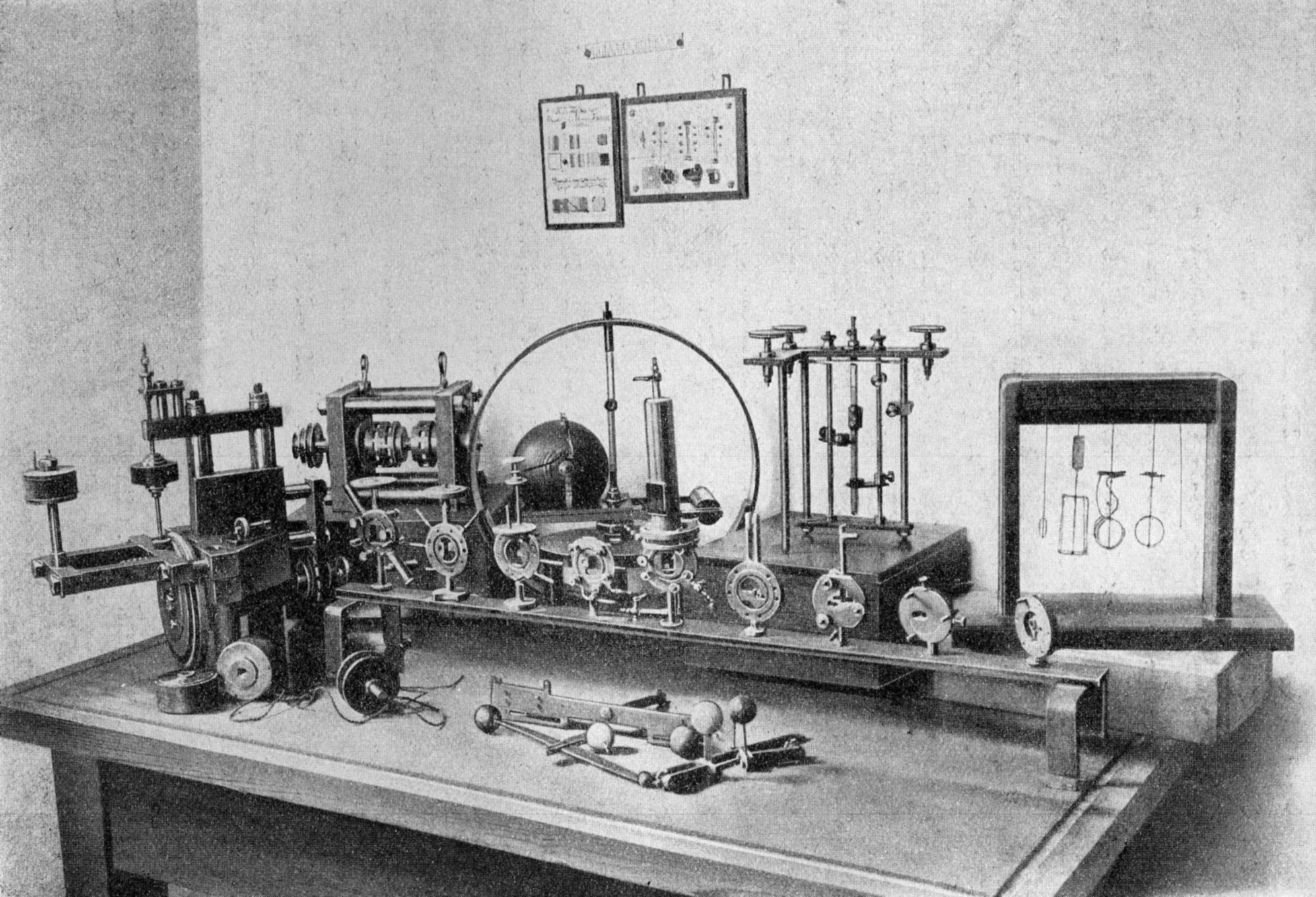}}
\vspace{1mm}
\centerline{\small P. N. Lebedev experimental equipment.}
~\\[-13mm]~
\end{figure}

Electromagnetic waves interact with matter. This interaction can be
explained as follows. We assume that one has a pendulum. The
pendulum is a massive body that can vibrate with a specific time
period or with a specific frequency $\omega$. If one exerts on the
pendulum a force that is just in resonance with the pendulum
vibrations, the pendulum starts to oscillate with increasing
amplitude. This means that the pendulum takes the energy from the
source which acts on the pendulum. Having this picture in mind,
P.~N.~Lebedev considered small objects including molecules as
analogs of such macroscopic pendulums. The electromagnetic waves
with frequencies corresponding to light of different colors provide
the energy of magnetic and electric field vibrations to the small
pieces of the macroscopic object. There is an analogy of the same
phenomenon if one considers the water waves on the surface of a
lake. If there is a boat on the lake, the waves on the lake reach
the boat, and the boat feels the action of the force associated with
the waves. The boat starts to move.

P.~N.~Lebedev and other physicists of this period of time (80$^{\rm
th}$--90$^{\rm th}$ years of the XIX Century) have a clear
understanding of the phy\-sical phenomena of the prequantum era. All
the vibrations, such as the acoustic vibrations in the air,
hydrodynamic vibrations, and electromagnetic vibrations, have common
properties. The difference is what vibrates when, e.g., acoustic
waves of the sound propagate in media or when electromagnetic waves
propagate in the vacuum or in media.

\begin{wrapfigure}[12]{r}{53mm}
\vspace{-6mm}
\centerline{\includegraphics[width=53mm]{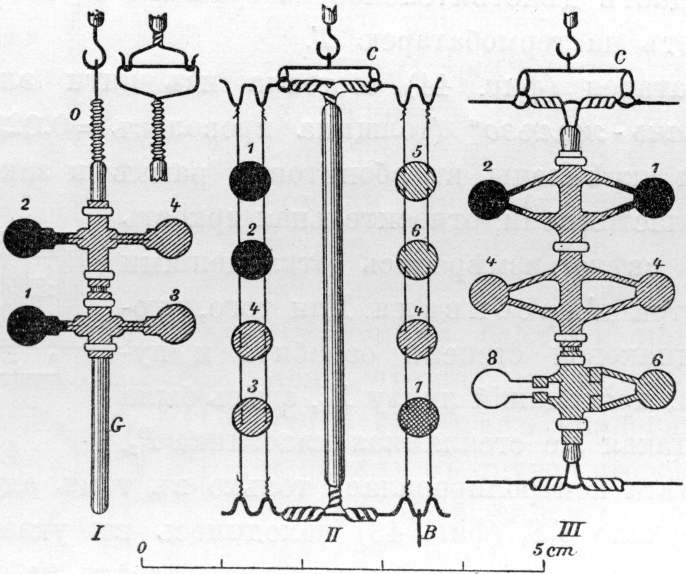}}
\vspace{1mm}
{\small Different wing models for Lebedev's experiments. }
\end{wrapfigure}

In the acoustic waves, the pressure and density of the air vibrate,
and these vibrations are connected due to the wave equation of the
sound. In the electromagnetic waves, the electric and magnetic
fields vibrate, and the vibrations are also described by the wave
equation, which directly follows from the Maxwell equations.

It is worth noting that, as is known, the Einstein equations of the
general relativity theory also provide the wave equation for small
perturbations of the gravitational field. The Einstein equations are
diffe\-rent from the Maxwell equations. Nevertheless, they have
solutions corresponding to the gravitational waves.
Recently~\cite{Science2016} gravitational waves acting on mirrors of
LIGO (Light Interferometer Gravitational-Wave Observatory) detector
were observed. This result is an experimental observation of the
action of gravitational radiation on a macroscopic object such as a
mirror.

This observation is a complete analog of the experiment performed by
P.~N.~Lebedev, where the action of the electromagnetic radiation on
a macroscopic object was observed in very sophisticated but
convincing experiment. In the discussed phenomena, the universal
properties of the vibrations and waves of arbitrary nature are
demonstrated. In fact, the sound, light, surface waves on water, and
the gravitational field vibrations seem to be (and they are)
absolutely different physical phenomena. But all of them have common
features and common behavior. The vibrations and waves carry energy,
and the waves act on the matter and macroscopic bodies, and this
action is analogous to the light pressure observed and measured by
P.~N.~Lebedev.

\section{Quantum Era}  
P.~N.~Lebedev passed away in 1912, being only 46 years old. His
scientific activity was dedicated to classical physics phenomena
since the quantum revolution in understanding the physical nature of
behavior of photons, electrons, and atoms was at the beginning of
the Twentieth Century in a very initial but fast developing stage.
The result of  this revolution was the formulation of quantum
mechanics, which could be associated with the introduction of the
notion of wave function and the finding by E.~Schr\"odinger of the
equation for the wave function in 1926~\cite{Schrod26}. The notion
of density matrix generalizing the wave function to the case of
thermal fluctuations was introduced in \cite{Landau27}; the
scientific spirit of this time in Russia is shown in \cite{Bereza}.

\begin{figure}[h]
\vspace{-1mm}
\centerline{\includegraphics[width=40mm]{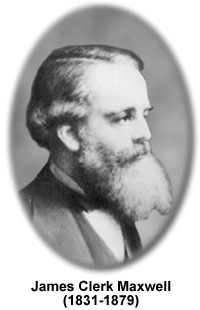}~
\includegraphics[width=40mm]{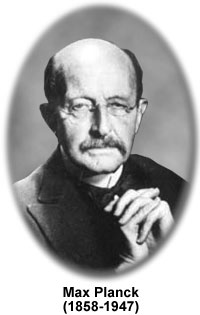}~
\includegraphics[width=40mm]{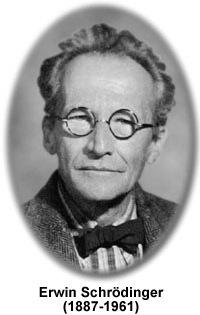}~
\includegraphics[width=40mm]{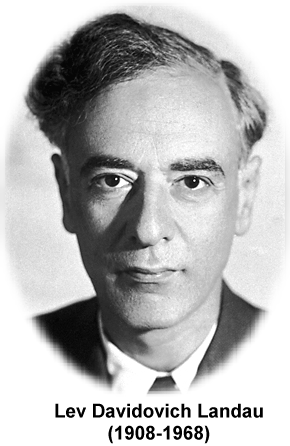}}
~\\[-9mm]~
\end{figure}

The Schr\"odinger equation and its applications to the description
of vibrations of any kind, including the electromagnetic field
vibrations, and the light pressure phenomenon observed by
P.~N.~Lebedev provided a new vision of all physical processes in
nature. Today, we live in the quantum era, and the behavior of
elementary particles like electrons and protons, atoms and
molecules, as well as the understanding of the creation and
evolution of the Universe, biological processes, chemical reactions,
in fact, all the natural processes one can imagine, are explained
and interpreted within the framework of quantum laws.

The quantum behavior of electromagnetic vibrations, in particular,
with frequencies corresponding to the visible-light electromagnetic
waves, added new aspects to the light pressure observed by
P.~N.~Lebedev. The set of theoretical and experimental researches of
the phenomena, which today are called optomechanical phenomena,
studied by optomechanics appeared (see, e.g.,~\cite{Tombesi} and
references therein). Ponderomotive forces were studied in
optomechanical phenomena by S.~Mancini, V.~I.~Man'ko, and
P.~Tombesi~\cite{Miranovic}.

\begin{wrapfigure}[9]{r}{100mm}
\vspace{-4mm}
\centerline{\includegraphics[height=38mm]{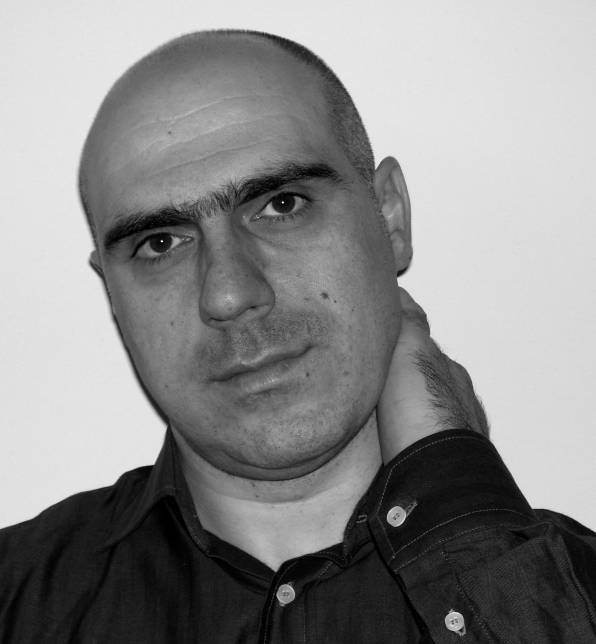}
\includegraphics[height=38mm]{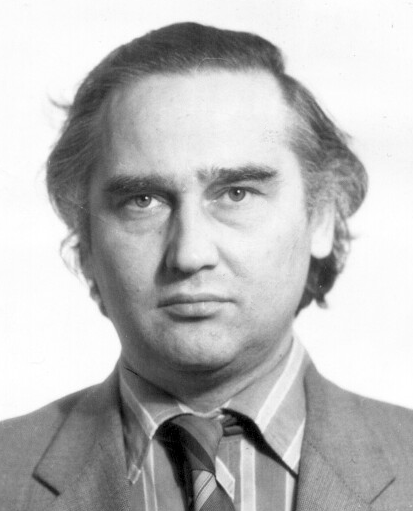}
\includegraphics[height=38mm]{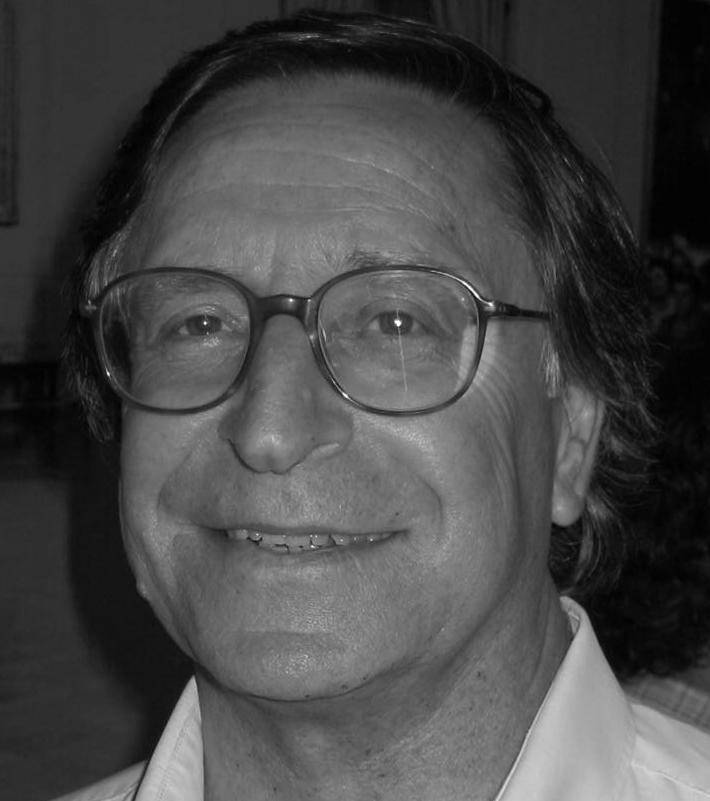}}
\centerline{\small S.~Mancini, V.~I.~Man'ko, and P.~Tombesi }
\end{wrapfigure}

In fact, it is a development of the Lebedev research on the
interaction of the light photons with macroscopic objects. In
optomechanics, the photons interact with oscillating macroscopic
object (mirror), but the oscillations of the electric and magnetic
fields in the electromagnetic wave and oscillations of the mirror
are studied using the quantum picture of the vibrations described by
the Schr\"odinger equation.

\begin{wrapfigure}[13]{l}{40mm}
\vspace{-1mm}
\centerline{\includegraphics[height=56mm]{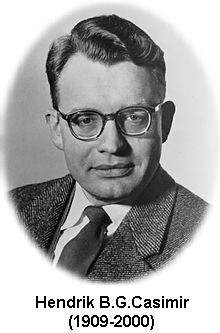}}
\end{wrapfigure}

The classical Maxwell equations do not contain the Planck constant
$h=6.62607004\cdot10^{-34}$~m$^2\cdot\,$kg/s
($\hbar=1.05457173\cdot10^{-34}$~m$^2\cdot\,$kg/s), which is
responsible for the explanation of quantum aspects of the
vibrations. The formulas for the light pressure used by
P.~N.~Lebedev and following from the classical equations for the
phenomenon of light interaction with matter also do not contain the
Planck constant. The formulas of optomechanics take into account the
presence of the Planck constant as well as the formulas of quantum
optics generalizing the formulas of classic optics. Below we discuss
one of the most impressive effects of quantum nature, called the
nonstationary Casimir effect, suggested at the Lebedev Physical
Institute by V.~V.~Dodonov, O.~V.~Man'ko, and
V.~I.~Man'ko~\cite{JRLR1989,VI-JRLR90,Olga-Korean}, or the dynamical
Casimir effect suggested by J.~Schwinger~\cite{Schwinger}; see the
history of the nonstationary Casimir effect in the
review~\cite{DodDodJRLR}.

\begin{wrapfigure}[11]{r}{106mm}
\vspace{-3mm}
\centerline{\includegraphics[height=42mm]{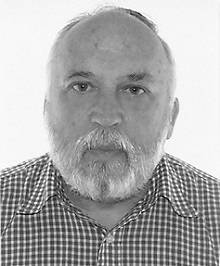}
\includegraphics[height=42mm]{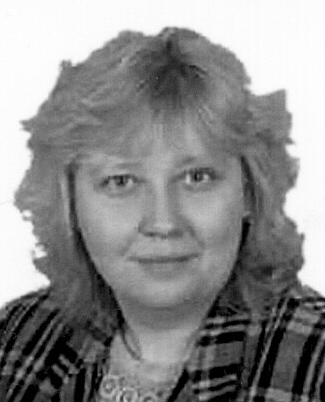}
\includegraphics[height=42mm]{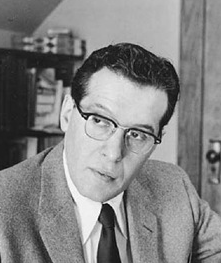}}
\centerline{\small V.~V.~Dodonov, O.~V.~Man'ko, and J.~Schwinger}
\end{wrapfigure}

To understand the reason for the effect, which means the creation of
photons by the action of two vibrating mirrors on the ``vacuum''
between the mirrors, we recall the properties of a quantum
oscillator or quantum pendulum. The usual classical pendulum can be
in the state of complete rest where the position and velocity of the
pendulum are equal to zero. The quantum nature of the pendulum
vibrations reflected by the Schr\"odinger equation describing all
possible states of this device prohibits such states of complete
rest. Such states contradict the uncertainty relation of the
pendulum position and momentum discovered by
Heisenberg~\cite{Heisenberg1926} and in generalized form by
Robertson~\cite{Robertson1926} and
Schr\"odinger~\cite{Schrodinger1930}.

The Heisenberg uncertainty relation
$$(\delta x)^2(\delta p)^2\geq\hbar^2/4,$$
where $\hbar$ is the Planck constant and $(\delta x)^2$ and $(\delta p)^2$ are
variances of the pendulum position and momentum, respectively, shows that one
cannot ``stop'' the pendulum in its state of rest. The pendulum positions
always fluctuate, and the degree of fluctuations has the bound determined by
the Planck constant $\hbar$. This constant is so small that, for the usual
macroscopic objects and in the experiments with classical devices of the type
employed by P.~N.~Lebedev, one cannot detect these fluctuations. Nevertheless,
in modern experiments with high-precision devices the influence of quantum
fluctuations of the pendulum position and momentum can be detected. We speak
of a pendulum, but the quantum uncertainty relation and the Schr\"odinger
equation, which automatically take into account the bound for the product of
the momentum-and-position variances determined by the Planck constant, are
valid for vibrations of any nature.

\begin{wrapfigure}[12]{r}{37mm}
\vspace{-7mm}
\centerline{
\includegraphics[height=58mm]{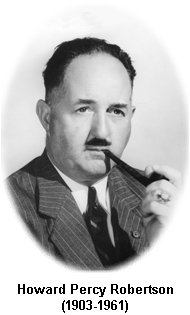}
}\end{wrapfigure}

P.~N.~Lebedev in his experiments studied the electromagnetic field
vibrations and vibrations in the waves propagating in the water; he
pointed out the universality of the properties of vibrations of
arbitrary kinds. His observation of the universality of vibration
properties turned out to be true not only in classical physics but
also in the quantum regime of vibrations unknown at the time of
Lebedev's scientific activity.

\begin{wrapfigure}[13]{l}{37mm}
\vspace{-2mm}
\centerline{\includegraphics[height=58mm]{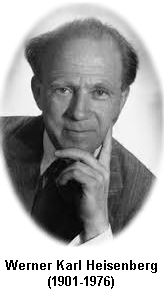}}
\end{wrapfigure}

The consequence of the existence of unavoidable quantum fluctuations of the
electromagnetic field vibrations is the existence of the Casimir attractive
force~\cite{Casimir48} between two parallel mirrors. If there is a complete
vacuum (no photons) between the mirrors, the mirrors attract with a force
proportional to the Planck constant $\hbar$. This kind of attraction is not
explained by the classical Newton and Maxwell equations, which do not contain
the Planck constant; this attraction is a completely quantum effect.

If varying in time one changes the distance between the mirrors
acting against the Casimir force, the work produced by such change,
due to the energy conservation law, has the effect of transforming
the vacuum (no photons or a state with mean photon number equal to
zero) to the state with mean value of photons different from zero.
This means that we have the generation of photons from the vacuum;
this effect was called the nonstationary~\cite{JRLR1989} or
dynamical~\cite{Schwinger} Casimir effect, and it was recently
observed in \cite{Sweden,Science2016}. It is worth noting that
photons created due to the nonstationary (dynamical) Casimir effect
are in squeezed states minimizing the Schr\"odinger--Robertson
uncertainty relations; this property of the created photons was
predicted in \cite{Klimov}.

The dynamical Casimir effect can be interpreted as the influence of
negative light pressure on the mirrors, which constitute the
resonator for the electromagnetic field oscillations. The attractive
Casimir force of two mirrors means that the pressure in the photon
``vacuum'' in the resonator is opposite to the light pressure that
repulses the mirrors. The joint effect of the action of external
forces changing the distance between the mirrors due to the work
produced by the outer sources and the Casimir forces is just the
generation of light photons by the light vacuum generator. Thus, the
positive light pressure measured by P.~N.~Lebedev in the quantum
domain of the electromagnetic field vibrations in the state, where
the mean number of photons between two parallel plane mirrors is
equal to zero, becomes negative.

It is an interesting and unexpected quantum phenomenon that does not exist in
classical physics. Also it is interesting that such light pressure in the case
of another mirror geometry or another form of the resonator, e.g., spherical
resonator, can be positive~\cite{Boyer}. The other kind of vibrations is
present in electric circuits constructed by the capacitance $C$ and inductance
$L$ connected by a wire. In such electric circuits, the current and voltage
vibrate exactly in a way as the position and velocity of the mechanical
pendulum do. The frequency of these vibrations $\omega$ is determined by the
circuit inductance and capacitance as $\omega\sim(LC)^{-1/2}$. If these
parameters are so small that the energy quantum of the vibration $\hbar\omega$
becomes larger than the thermal energy $k_BT$, all quantum properties of the
vibrations start to be visible and detectable.

First of all, quantum fluctuations of the current and voltage in the
circuit (quantum noise) satisfy the Heisenberg and
Schr\"odinger--Robertson uncertainty relations. The current--voltage
characteristics of the electric circuit cannot be measured, since
the current and voltage in the quantum regime of vibrations cannot
be measured simultaneously with high precision. This is impossible
not because there is no device in electrotechnics that can measure
the current and voltage with high precision, but because such
devices cannot exist, in principle, due to the quantum nature of any
kind of vibrations, including the vibrations of the current and
voltage in the electric circuit, which should satisfy the Heisenberg
uncertainty relation.

Analogously, for a mechanical pendulum, in principle, it is
impossible to measure simultaneously the pendulum position and its
velocity with very high precision, since these characteristics
fluctuate. The vibrations in electric circuits, if the capacitance
or the inductance (or both) change in time, behave as the
electromagnetic field vibrations in the resonator between two
parallel mirrors forming this resonator. There was a
suggestion~\cite{JRLR1989,VI-JRLR90,Olga-Korean} to employ such an
electric circuit realized by the superconducting circuit, which is a
device based on the properties of the Josephson junction, to
generate the current in this circuit by varying the junction
parameters in time.

The generation of current in a circuit with varying capacitance and
inductance is completely analogous to the generation of photons in a
resonator with time-varying boundaries observed in
\cite{Science2016}. This analogy follows from the universality of
quantum properties of any kind of vibrations; such properties of the
nonstationary Casimir effect were recently studied in
\cite{Zeilinger1,Zeilinger2}.

The vibrations in liquids and solids provide the phenomenon of sound
propagation. In the quantum picture, the sound is treated as
phonons, which are analogs of photons in the case of
electromagnetic-wave propagation, e.g., light propagation. In view
of the universality of quantum properties of vibrations, the phonons
and the sound waves should behave in specific conditions analogously
to the behavior of photons and electromagnetic waves.

\begin{wrapfigure}[13]{l}{46mm}
\vspace{-1mm}
\centerline{\includegraphics[height=58mm]{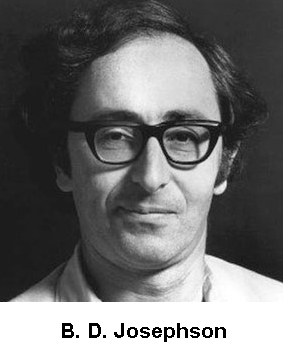}}
\end{wrapfigure}

For example, an analogy of the Casimir effect and nonstationary Casimir effect
should exist; this means, in particular, that generation of the sound quanta
(phonons) can exist in sound resonators with moving boundaries. Such
possibility was discussed in sound in liquid helium, e.g., in \cite{Volovik}.
%\cite{JRLR1989,VI-JRLR90,Olga-Korean}.
Since the gravitational wave is the other phenomenon with vibrations of
physical characteristics of the gravitational field, in view of the
universality of quantum behavior of vibrations, the possibility of existence
of a gravitational analog of the Casimir effect and nonstationary
gravitational Casimir effect with generation of gravitons from the vacuum
state can be conjectured (see, e.g.,~\cite{Ruser}). Thus, a quantum generator
of gravitons or the quantum sound gene\-rator, in principle, can be thought
off as possible devices of future technology.

Quantum technologies associated with the development of quantum
information and quantum computing based on applications of
superconducting circuits based on Josephson junctions to realize
qubits are studied, for example, in
\cite{Kiktenko1,Kiktenko2,GlushkovJRLR}.

\newpage
\section{Lebedev Institute}
P. N. Lebedev was not only the experimentalist who measured in the
laboratory the pressure of light; he also had a strong group of
young colleagues and students who became famous scientists and
developed further research in Russia. S.~I.~Vavilov was one of the
representatives of the Lebedev group. During the time of Lebedev's
scientific activity and later on, his laboratory was converted to
the famous Physical Institute of the USSR Academy of Sciences named
in 1934 {\it the P.~N.~Lebedev Physical Institute}.

\begin{figure}[ht]
\vspace{-1mm}
\centerline{\includegraphics[height=56mm]{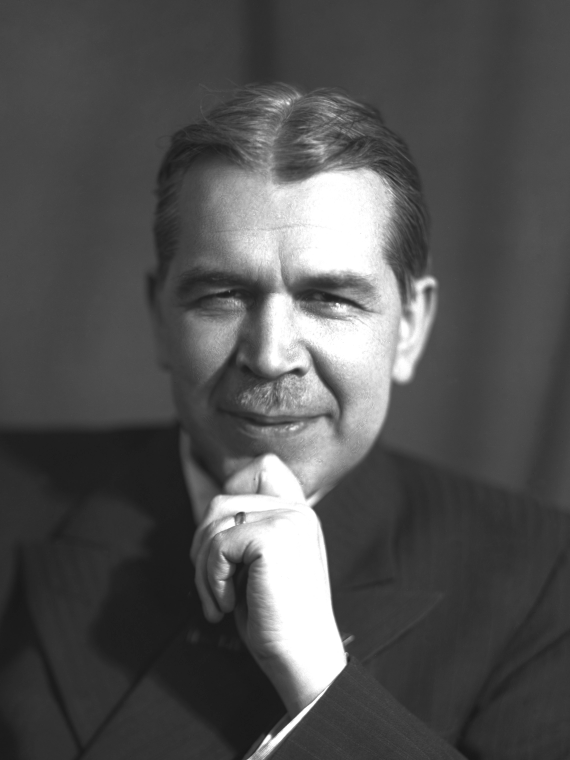}\,
\includegraphics[height=56mm]{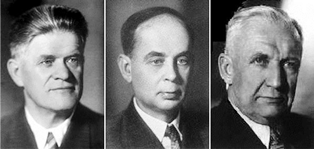}}
\centerline{\small S.~I.~Vavilov, P.~A.~Cherenkov, I.~M.~Frank, and I.~E.~Tamm}
~\\[-11mm]~
\end{figure}

\begin{wrapfigure}[16]{r}{88mm}
\vspace{-5mm}
\centerline{\includegraphics[width=87mm]{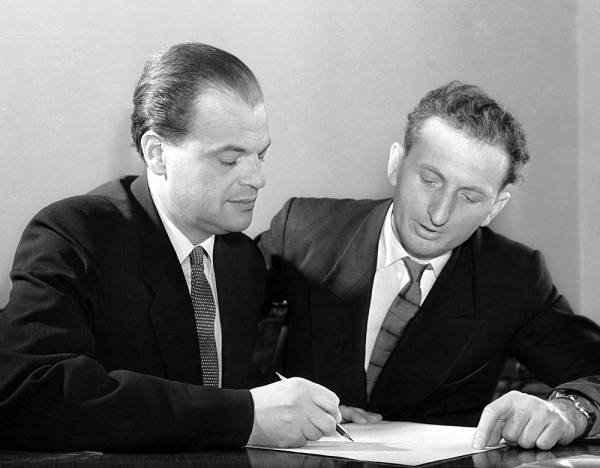}}
\vspace{1mm}
\centerline{\small  N.~G.~Basov and A.~M.~Prokhorov}
\end{wrapfigure}

Academician S.~I.~Vavilov initiated the investigation of a charged
particle moving in media with constant velocity larger that the
light velo\-city in the media. P.~A.~Cherenkov, PhD student of
S.~I.~Vavilov observed the radiation created by such moving charged
particle; this radiation was called the Cherenkov radiation or
Vavilov--Cherenkov radiation. A rigorous theory of the
Vavilov--Cherenkov radiation was developed by I.~M.~Frank and
I.~E.~Tamm, professors of the Lebedev Institute.

The experimental and theoretical researches performed in the
P.~N.~Lebedev Physical Institute by A.~M.~Prokhorov and N.~G.~Basov
in the 60th years of the last century created a new area of physics
and technology based on the quantum properties of photons
interacting with atomic and molecular vibrations. Namely, masers and
lasers were created due to the fundamental investigations of
universal properties of vibrations of any kind discussed in
classical physics by P.~N.~Lebedev but also existing in the quantum
domain of the behavior of matter. It is worth mentioning that, due
to a mysterious coincidence, the laboratory of the P.~N.~Lebedev
Physical Institute where the pioneer works on the creation of masers
and lasers were performed by the groups of A.~M.~Prokhorov and
N.~G.~Basov was called {\it The Laboratory of Vibrations
(Oscillations)}.

\section*{Acknowledgments}  
We are grateful to Prof. V. I. Man'ko for valuable comments.

\end{document}